\documentclass[aps,pra,twocolumn,showpacs,floatfix,groupedaddress]{revtex4-1}
\usepackage{eurosym}
\usepackage{graphicx}
\usepackage{subfigure}
\usepackage{color}
\usepackage{soul}
\usepackage{textcomp}

\begin{document}

\title{Reflective Optical Limiter Based on Resonant Transmission}
\author{Eleana Makri, Tsampikos Kottos}
\address{Department of Physics, Wesleyan University, Middletown CT-06459, USA}

\author{Ilya Vitebskiy}
\address{The Air Force Research Laboratory, Sensors Directorate, Wright Patterson AFB, OH 45433 USA}

\begin{abstract}
Optical limiters transmit low-level radiation while blocking electromagnetic
pulses with excessively high energy (energy limiters) or with excessively
high peak intensity (power limiters). A typical optical limiter absorbs most
of the high-level radiation which can cause its destruction via overheating.
Here we introduce the novel concept of a reflective energy limiter which
blocks electromagnetic pulses with excessively high total energy by
reflecting them back to space, rather than absorbing them. The idea is to
use a defect layer with temperature dependent loss tangent embedded in a
low-loss photonic structure. The low energy pulses with central frequency
close to that of the localized defect mode will pass through. But if the
cumulative energy carried by the pulse exceeds certain level, the entire
photonic structure reflects the incident light (and does not absorb it!) for a
broad frequency window. The underlying physical mechanism is based on
self-regulated impedance mismatch which increases dramatically with the
cumulative energy carried by the pulse.
\end{abstract}

\maketitle

\address{Department of Physics, Wesleyan University, Middletown CT-06459,
USA}

\address{The Air Force Research Laboratory, Sensors Directorate, Wright
Patterson AFB, OH 45433 USA}


\section{Introduction}

The protection of photosensitive optical components from high incident
radiation has applications to areas as diverse as microwave and optical communications
to optical sensing \cite{ST91,O97,PHR98}. As a result, a
considerable research effort has focused on developing novel protection
schemes and materials that provide control of high-level optical and
microwave radiation and prevent damages of optical sensors (including the
human eye) and microwave antennas \cite{limiter1,limiter2,limiter3,
RC69,BSMBS85,DJ95}. Optical limiters constitute an important class of such
protection devices. They are supposed to transmit low-level radiation, while
blocking light pulses with high level of radiation. A typical
passive optical limiter absorbs most of the high-level radiation, which can
cause its destruction via overheating. The most common set-up of a passive
optical limiter consists of a single protective layer with complex
permittivity $\epsilon =\epsilon ^{\prime }+i\epsilon ^{\prime \prime }$,
where the imaginary part $\epsilon ^{\prime \prime }$ increases sharply with
the radiation level. For low-level radiation, the absorption is negligible,
and the protective layer is transparent. An increase in the radiation level
results in an increase in $\epsilon ^{\prime \prime }$, which renders the
protective layer opaque. As a consequence, most of the high-level radiation
will be absorbed by the limiter, which can cause its overheating and
destruction. It turns out that if the same protective layer is incorporated
into a certain photonic layered structure, the entire multilayer can become
highly reflective for high-level radiation, while remaining transmissive at
certain frequencies if the radiation level is low. Such a reflective limiter
can be immune to overheating and destruction by high-level laser radiation,
which is our main objective.

The physical reasons for the sharp increase in $\epsilon ^{\prime \prime }$
with the radiation level can be different. For instance, it can be photoconductivity, heating,
two-photon absorption, or any combination of the
these mechanisms. In our previous publication \cite{MRKV14} we considered
the particular case of a strong non-linear dependence of $\epsilon ^{\prime
\prime }$ of the protective layer on light intensity. This can be
attributed, for instance, to a two-photon absorption. We showed that
incorporation of such a nonlinear layer in a properly designed low-loss
layered structure makes the entire assembly act as a reflective power
limiter. In this paper, we consider a more practical particular case where
the increase in $\epsilon ^{\prime \prime }$ is due to heating of the
protective layer. We show that, depending on the pulse duration as compared to the
thermal relaxation time, the
properly design layered structure incorporating such a protective layer can
act as a reflective energy limiter, or as a reflective power limiter.
Specifically, for short pulses, such a layered structure acts as an energy limiter, reflecting light pulses carrying 
excessively high energy. By
comparison, for sufficiently long pulses, the 
same structure will
act as a power limiter. In either case, most of the incident radiation will
be reflected back to space, even though a stand-alone protective layer would
act as an absorptive optical limiter.

The proposed architecture consists of a (protective) defect layer embedded
in a low-loss Bragg grating . In contrast to the reflective power limiter
introduced in \cite{MRKV14}, the defect layer does not have to be nonlinear,
but it must display strong temperature dependence $\epsilon ^{\prime \prime}(T)$ of the imaginary part of its 
permittivity. If the total energy carried by the pulse is low, $\epsilon ^{\prime \prime}(T)$
remains small enough to support a localized mode and the resonant
transmittance associated with this mode. If, on the other hand, the energy
carried by the pulse exceeds certain level, the defect layer becomes lossy
enough to suppress the localized mode, along with the resonant
transmittance. The entire stack turns highly reflective, which is consistent
with our goal. We refer to this limiter as a \textit{reflective energy
limiter} in order to distinguish it from the nonlinear reflective power
limiter introduced in \cite{MRKV14}. Finally, if the pulse duration
significantly exceeds the thermal relaxation time of the defect layer, the
entire layered structure will again act as a reflective power limiter with
the cut-off light intensity determined by the thermal relaxation time of the
defect layer -- not by the nonlinearity in $\epsilon ^{\prime \prime }$, as
was the case in \cite{MRKV14}.

The organization of the paper is as follows. In Sec. \ref{model}, a conceptual design
for the reflective energy limiter is presented, along with the mathematical
formalism used in our calculations. In Sec. \ref{Theory}, we analyze the
role of thermal conductivity. The latter plays an important role if the
pulse duration is comparable or exceeds the thermal relaxation time of the
defect layer.

\begin{figure}[th]
\includegraphics[width=1\linewidth, angle=0]{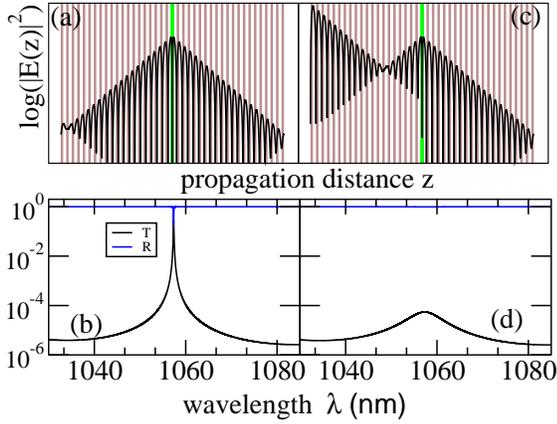}
\caption{(Color online) A schematics of a reflective energy limiter. Two identical lossless Bragg reflectors are placed
on the left and right of a lossy layer (green). The value of $\protect\epsilon ^{\prime \prime }$ in the defect layer is 
an increasing function of temperature. (a) Field distribution at the frequency of resonance transmission for an incident 
pulse with low energy   -- the field amplitude at the location of the defect layer is exponentially higher than that of the 
incident wave. (b) Transmittance vs. light wavelength for low incident light energy. (c) Field distribution at the frequency 
of maximum transmittance for an incident pulse with high energy --the amplitude of the suppressed localized mode 
is lower than that of the incident wave. (d) Transmittance vs. wavelength for an incident pulse with high energy.}
\label{fig1}
\end{figure}


\section{Physical Structure and mathematical model}\label{model}

We consider two identical losses Bragg reflectors consisting of two alternating
layers. Each mirror consists of forty layers which are placed at $-L\leq z
\leq 0$ and $d\leq z\leq L+d$. For the sake of the discussion we assume that
the layers consist of Al$_2$O$_3$ and SiO$_2$ with corresponding
permittivities $\epsilon_1=3.08$ and $\epsilon_2=2.1$. These values are
typical for these materials at wavelengths $\lambda \sim 1 \mu m$. The width
of layers is assumed to be $d_1=151nm$ and $d_2\approx 183nm$
respectively. At $0\leq z\leq d$ we introduce a defect lossy layer with
complex permittivity $\epsilon_{\mathrm{d}} = \epsilon_{\mathrm{d}}^{\prime
}+i\epsilon_{\mathrm{d}}^{\prime \prime }$. We further assume that the
imaginary part of the permittivity of the defect layer depends on the
temperature $T$ i.e. $\epsilon_{\mathrm{d}}^{\prime \prime }=\epsilon_{%
\mathrm{d}}^{\prime \prime }(T)$. For simplicity, we assume linear
dependence i.e. $\epsilon_{\mathrm{d}}^{\prime \prime }(T)=c_1+c_2 T$ where $%
c_1,c_2$ are some characteristic constants of the defect. Below we assume
that $\epsilon_{\mathrm{d}}^{\prime }=12.11$ (which is a typical value for,
say GaAs, at near infrared), $c_1=10^{-5}$ and $c_2= 1$ while the width of
the defect layer is taken to be $d=151nm$.

The transmittance $\mathcal{T}$, reflectance $\mathcal{R}$ and absorption $\mathcal{A}$ 
of our set up, and the
field profile at any frequency can be calculated via the transfer matrix
approach. Specifically, a monochromatic electric field of frequency $\omega$
satisfies the Helmholtz equation: 
\begin{equation}
{\frac{\partial ^{2}E(z)}{\partial z^{2}}}+{\frac{\omega ^{2}}{c^{2}}}\epsilon (z)E(z)=0\,\,\,.  
\label{Helm}
\end{equation}
At each layer inside the grating, Eq.~(\ref{Helm}) admits the solution $E^{(j)}= E_f^{(j)} \exp(in_j k z) + E_b^{(j)}\exp(-in_j k z)$, 
where $n_j=\sqrt{\epsilon_j}$ is the refraction index of the $j$-th layer and $k$ is the wave vector $k=\omega /n_{0}c$ ($c$ is 
the speed of light in the vacuum
and $n_0$ is the refractive index of air). Imposing continuity of the field
and its derivative at each layer interface, as well as taking into
consideration the free propagation in each layer, we get the following
iteration relation 
\begin{eqnarray}  \label{transfer0}
\left(\begin{array}{c}
E_{f}^{(j)} \\ 
E_{b}^{(j)}\end{array}
\right) & = & \mathcal{M}^{(j)} \left(\begin{array}{c}
E_{f}^{(j-1)} \\ 
E_{b}^{(j-1)}\end{array}
\right); \mathcal{M}^{(j)}= P_{R}^{(j)} Q^{(j)} K^{(j)} P_{L}^{(j)}.
\end{eqnarray}
where 
\begin{eqnarray}  \label{sltm}
Q^{(j)}=&\left(\begin{array}{cc}
e^{ikn_jd_j} & 0 \\ 
0 & e^{-ikn_jd_j}
\end{array}
\right)  \nonumber \\
K^{(j)}=& \left(\begin{array}{cc}
{\frac{n_j+n_{j-1}}{n_j}} & {\frac{n_j-n_{j-1}}{n_j}} \\ 
{\frac{n_j-n_{j-1}}{n_j}} & {\frac{n_j+n_{j-1}}{n_j}}
\end{array}
\right)  \nonumber \\
P_{R}^{(j)}=&\left(
\begin{array}{cc}
e^{ikn_jz} & 0 \\ 
0 & e^{-ikn_jz}
\end{array}
\right)  \nonumber \\
P_{L}^{(j)}=&\left(\begin{array}{cc}
e^{-ikn_{j-1}(z-d_j)} & 0 \\ 
0 & e^{ikn_{j-1}(z-d_j)}\end{array}
\right)
\end{eqnarray}

At the same time the field outside the layered structured can be written as $E_{0}^{-}(z)=E_{f}^{-}
\exp(ikz)+E_{b}^{-}\exp (-ikz)$ for $z<-L$ and $E_{0}^{+}(z)=E_{f}^{+}\exp (ikz)+E_{b}^{+}\exp (-ikz)$ 
for $z>L+d$. The amplitudes of forward and backward propagating waves on the left $z<-L$ and
right $z>L+d$ domains are related via the total transfer matrix $\mathcal{M}
=P_{R}^{(2N+2)} K^{(2N+2)}\Pi_j \mathcal{M}^{(j)}$ (where $N$ is the number of layers on each grating and 
$n_{2N+2}=n_{0}$): 
\begin{eqnarray}  
\label{transfer}
\left(\begin{array}{c}
E_{f}^+ \\ 
E_{b}^+\end{array}
\right) & = & \left(
\begin{array}{cc}
\mathcal{M}_{11} & \mathcal{M}_{12} \\ 
\mathcal{M}_{21} & \mathcal{M}_{22}
\end{array}
\right) \left(
\begin{array}{c}
E_{f}^- \\ 
E_{b}^-
\end{array}
\right)
\end{eqnarray}
The transmission and reflection coefficients and the field profile, say for
a left incident wave, can be obtained by iterating backwards Eqs.~(\ref{transfer0},\ref{transfer}) together 
with the boundary conditions $E_b^+=0$ and $\left|E_f^+\right|=1$ (due to the linearity of the equations, one can
always impose a value for the outgoing field and calculate via a backward
iteration of the transfer matrices the input field to which corresponds 
\cite{backward}). Specifically we have $\mathcal{T}\equiv|E_{f}^{+}/E_{f}^{-}|^{2}$; $\mathcal{R}\equiv
|E_{b}^{-}/E_{f}^{-}|^{2}$. These can be expressed in terms of the transfer
matrix elements as $\mathcal{T}= \left|{\frac{1}{\mathcal{M}_{22}}}\right|^2; 
\mathcal{R}= \left|{\frac{\mathcal{M}_{21}}{\mathcal{M}_{22}}}\right|^2$.
The absorption coefficient $\mathcal{A}$ can then be evaluated in terms of
transmittances and reflectances as $\mathcal{A}\equiv 1- \mathcal{T} -\mathcal{R}$.

\section{Theoretical analysis}

\label{Theory} In the case that the permittivity of the defect layer is replaced by
$\epsilon _{d}=\epsilon _{1}$, the whole
structure is periodic and displays a typical dispersion relation consisting
of transparent frequency windows (bands) where light is transmitted with
near-unity transmittance alternated with frequency windows (gaps) where the
incident light is experiencing almost complete reflection.

When the defect is included in the middle of the grating, for zero
temperature $T=0$ corresponding to permittivity $\epsilon_{ \mathrm{d}}\approx 
\epsilon_{\mathrm{d}}^{\prime }$, the layered structure supports a
localized resonant defect mode (see Fig. \ref{fig1}a) with a frequency lying
in a photonic band gap of the Bragg grating (see Figs. \ref{fig1}b). For the
specific set up that we consider here, we find that a resonant mode is
located in the middle of the gap at wavelength $\lambda_r\approx 1060nm$.
This defect mode is localized in the vicinity of the defect layer and decays
exponentially away from the defect (see Fig. \ref{fig1}a). In the vicinity
of the localized mode frequency $\omega_r$, the entire layered structure
displays a strong resonant transmission due to the excitation of the
localized mode (see Fig. \ref{fig1}b). In other words, the transmittance is $\mathcal{T} \approx 1$ 
while the reflectance and the absorption in the
absence of any losses are $\mathcal{R}\approx 0$ and $\mathcal{A}\approx 0$
respectively. This picture is still applicable even in the presence of small
(but non-zero) dissipative permittivity $\epsilon_{\mathrm{d}}^{\prime
\prime} \neq 0$ (see Fig. \ref{fig1}a,b).

An alternative expression for the absorption coefficient $\mathcal{A}$ can be given in terms of the permittivity and the field intensity $|E(z)|^2$
inside the defect layer. The resulting expression is derived by subtracting the product of Eq. (\ref{Helm}) with $E^*(z)$ from its complex conjugate
form and then integrating the outcome over the interval $-L\leq z\leq L$. We get 
\begin{equation}  \label{A_wf}
\left(E^*{\frac{dE}{dz}} - E {\frac{dE^*}{dz}}\right)_{z=-L}^{z=L} +2 i k^2\int_{-L}^{L} \mathcal{I}m \epsilon(z) \left|E(z)\right|^2 dz =0.
\end{equation}
Substituting in Eq. (\ref{A_wf}) the expressions of the electric field at $z=-L$ and $z=L$ respectively we get 
\begin{equation}  
\label{Asfin}
\mathcal{A}\equiv 1-\mathcal{T} -\mathcal{R} = {\frac{k}{\left|E_f^-\right|^2}} \int_{-L}^{L} dz |E(z)|^2 \mathcal{I}m \epsilon(z).
\end{equation}
Furthermore\textcolor{red}{,} we assume that $\mathcal{I}m \epsilon (z)$ is zero everywhere inside the layered structure apart from the interval 
$0\leq z\leq d$ where the defect layer is placed. In this interval it takes a uniform value $\mathcal{I}m \epsilon (0\leq z\leq d) =
\epsilon_{\mathrm{d}}^{\prime \prime }(T)$. These simplifications allow us to express the absorption coefficient of Eq. (\ref{Asfin}) in the following 
form 
\begin{equation}
\mathcal{A}(T)=\rho\left( T\right) \omega\epsilon_{\mathrm{d}
}^{\prime\prime}\left( T\right)  \label{Afin}
\end{equation}
where $\rho\left( T\right) =\mathcal{I}_{\mathrm{d}}/\left|E_f^{-}\right|^2$ is the ratio of the integral of light intensity $\mathcal{I}_{\mathrm{d}}
=\int_0^d dz \left|E(z)\right|^{2}$ at the lossy layer and the incident light intensity. It is obvious from Eq. (\ref{Afin}) that $\mathcal{A}(T)$
depends on both the dissipative part of the permittivity and the value of the electric field inside the defect layer. Although the former increases
monotonically with the temperature $T$ and thus with the duration time of the incident pulse, this is not true for $\rho(T)$. The latter, which is a
unique function of the permittivity, remains approximately constant up to some value of $\epsilon_{\mathrm{d}}^{\prime \prime }$ above which it
decreases, leading eventually to a total decrease of the absorption coefficient together with a simultaneous increase of the reflectivity of the
structure. This is related to the fact that the increase of $\epsilon_{\mathrm{d}}^{\prime \prime }$ spoils the resonant localized mode (see Fig. 
\ref{fig1}c) which is responsible for high transmittance. Specifically, when the losses due to $\epsilon_{\mathrm{d}}^{\prime \prime }$ overrun the
losses due to leakage from the boundaries of the structure, the resonant mode cease to exist (see Fig. \ref{fig1}c) and the structure becomes
reflective, i.e. $\mathcal{R} \approx 1$, and $\mathcal{T} \approx 0$, see Fig. \ref{fig1}d. As a consequence we have that $\mathcal{A}=1-\mathcal{T}-
\mathcal{R}\approx 0$ and the system does not absorb the high incident energy of the incoming light source but rather reflects it back in space.

In fact, the non-monotonic shape of the envelope of the scattering field in Fig. \ref{fig1}c is a direct consequence of the fact that the structure
becomes reflective $\mathcal{R}\approx 1; \mathcal{T}\approx 0$. One has to realize that in the case that both Bragg gratings on the left and 
right of the defect layer are finite, the field inside each half-space is written as a linear combination of two evanescent contributions with 
exponentially decreasing and exponentially increasing amplitudes. Their relative weight is determined by the boundary conditions $E(z=-L) =
E_0^-(-L)$ and $E(L)=E_0^{+}(L)= E_f^- \sqrt{\mathcal{T}}$ at the two outer interfaces of the layered structure. In the case of reflective structures 
these boundary conditions lead to the relation $E(-L)=E_f^-\sim \mathcal{O}(1)$ and $E(L)\approx 0$. It can be shown rigorously that in this case, 
the field on the left half-space of the structure is dominated originally by the exponentially decaying component while after some turning point 
$z_0$ the exponentially increasing component becomes dominant up to the defect layer. After that the field decays exponentially as in the resonant 
case. Similar scattering field profiles have been found in cases of active (gain) defects \cite{PACY10}.

One can use a simple qualitative argument that allows to estimate the condition under which $\mathcal{A}(T)$ continues to increase. 
As we discuss previously, we assume that the electromagnetic energy losses occur in the lossy defect layer. The dissipated power can 
be estimated from Eq. (\ref{Afin}) to be ${\dot Q}\propto \mathcal{A} \cdot \left|E_f^-\right|^2=\omega \epsilon_{\mathrm{d}}^{\prime 
\prime }\mathcal{I}_{\mathrm{d}}$. Due to the energy conservation, the rate of energy dissipation cannot exceed the energy supply 
provided by the incident wave. The latter is $\mathcal{S}_{\mathrm{in}}\propto c \cdot \left|E_f^{-}\right|^2$. Taking this constraint
into account we get the following upper limit on the field intensity at the defect layer location 
\begin{equation}  \label{ineq}
{\frac{c}{\omega \epsilon^{\prime \prime }(T) d}} \left|E_f^{-}\right|^2\geq
\left|E_{\mathrm{d}}\right|^2
\end{equation}
Above we have made the additional approximation that $\mathcal{I}_{\mathrm{d}}\sim \left|E_{\mathrm{d}}\right|^2 \cdot d$, where 
$E_{\mathrm{d}}$ is a typical value of the field inside the defect layer.

Next we recall that a resonant mode with a frequency $\omega$ inside the band-gap has a Bloch wave number which is imaginary $k=
ik^{\prime \prime }$. The electric field inside the layered structure, can be expressed as a pair of evanescent modes, one of which is 
decaying with the distance $z$ and another one which is growing i.e. $E(z) = E_f \exp(-k^{\prime \prime }z) +E_b \exp(k^{\prime 
\prime }z)$. To the left of the defect ($-L<z<0$), the electric field is dominated by the rising evanescent mode $E(z)\approx E_b
\exp(k^{\prime \prime }z)$ while to the right of the defect ($0<z<L$), the dominant contribution is provided by the decaying mode 
$E(z)\approx E_f \exp(-k^{\prime \prime }z)$ \cite{notefield}.

The field $E_{\mathrm{d}}$ at the location of the defect layer is provided by the rising evanescent mode evaluated at $z=0$ i.e. 
$E_{\mathrm{d}}\sim E_b $. Therefore, the value of this evanescent mode at the left stack boundary at $z=-L$ is 
\begin{equation}  \label{equal}
E(-L) \propto E_{\mathrm{d}} \exp(-k^{\prime \prime }L)
\end{equation}
Comparing (\ref{ineq}) and (\ref{equal}) we can conclude that if 
\begin{equation}  \label{ineq2}
{\frac{c}{\omega \epsilon^{\prime \prime }(T) d}} \exp(-2k^{\prime \prime}L)\ll 1
\end{equation}
then the amplitude of rising evanescent mode $E(z=-L)$ at the left stack boundary is much less than amplitude of the incident wave 
\begin{equation}  
\label{inequal}
\left|E(z=-L)\right|^2\ll \left|E_f^{-}\right|^2.
\end{equation}
The latter condition Eq. (\ref{inequal}), implies that the energy density inside the left grating is much smaller than the energy density 
of the incident wave, hence, only a small portion of the incident light energy $S_I\propto c \left|E_f^{-}\right|^2$ will cross the stack 
boundary at $z=-L$. In other word, the condition Eq. (\ref{ineq2}) automatically implies high reflectivity at the stack interface. The 
condition Eq. (\ref{ineq2}) for high stack reflectivity (and hence low transmittance and absorption) will always be satisfied if the loss 
tangent $\epsilon^{\prime \prime }(T)$ of the defect layer is large enough and/or if the number of layers in the Bragg grating is large 
enough.

\begin{figure}[tbp]
\begin{center}
\includegraphics[width=1\linewidth, angle=0]{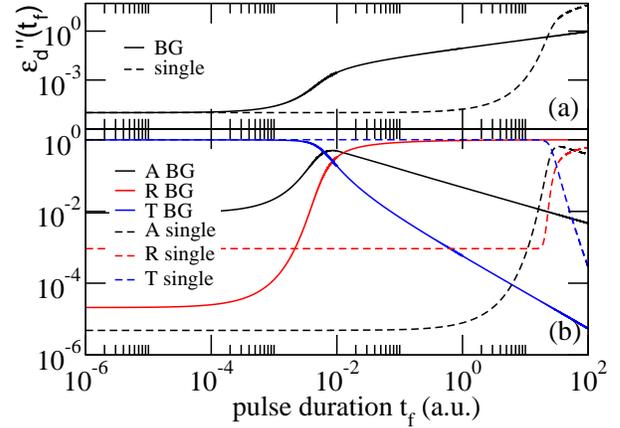}
\end{center}
\caption{(a) The imaginary part $\protect\epsilon _{d}^{\prime \prime }$ of permittivity as a function of pulse duration $t_{f}$. The solid 
line corresponds to the layered structure in Fig. \protect\ref{fig1}, while the dashed line corresponds to the stand-alone lossy layer. (b) 
The absorption $\mathcal{A}(t_{f})$ (black solid line), reflectance $\mathcal{R}(t_{f})$ (red solid lines) and transmission $\mathcal{T}(t_{f})$ 
(blue solid line) of the layered structure in Fig. \protect\ref{fig1} vs. pulse duration. For longer pulse duration (and larger cumulative energy 
of the pulse), the absorption $\mathcal{A}$ is suppressed and the set-up becomes highly reflective ($\mathcal{R}\approx 1$). The dashed 
lines show the respective values for the stand-alone lossy layer, in which case, the absorption for pulses with longer duration  (and larger 
cumulative energy)  is much higher, while the reflectivity is much lower than those of the layered structure in Fig. \protect\ref{fig1}.}
\label{fig2}
\end{figure}

\begin{figure}[tbp]
\begin{center}
\includegraphics[width=1\linewidth, angle=0]{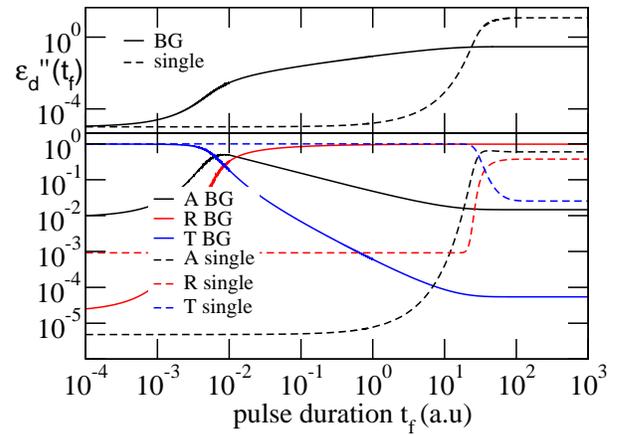}
\end{center}
\caption{(Color online) The same as in Fig. \protect\ref{fig2} but now in the presence of thermal exchange between the defect layer and 
its surroundings ($\protect\kappa =0.05$). For longer pulse duration, a steady state regime is reached, which corresponds to a crossover 
from energy limiting regime to a power limiting regime.}
\label{fig3}
\end{figure}

Next, we want to quantify the above arguments. To this end, we calculate explicitly the transport characteristics of our grating structure 
for an incident laser pulse. Although the analysis can be generalized for any incident pulse shape, in our numerical simulations below, 
we have assumed for simplicity that the incident laser pulse has a train-form \cite{notetrain} 
\begin{equation}
\begin{array}{cccc}
\mathcal{W}_I(t) & = 0 & \mathrm{for} & t\leq 0 \\ 
& =w_0 & \mathrm{for} & 0\leq t \leq t_f \\ 
& = 0 & \mathrm{for} & t\geq 0\nonumber
\end{array}
\end{equation}
We want to calculate the total energy transmitted, reflected, and absorbed during the duration of the pulse. These can be expressed in 
terms of the time-dependent transmittance $\mathcal{T}(t)$, reflectance $\mathcal{R}(t)$ and absorption $\mathcal{A}(t)$ which are 
the main quantities that we analyze below. All other observables can be easily deduced from them. For example, the integrated (over the 
period of the pulse) absorption $\bar {\mathcal{A}}$ can be defined as 
\begin{equation}
\bar{\mathcal{A}}= {\frac{\int\limits_{-\infty}^{\infty} dt\mathcal{A}(t)
\mathcal{W}_{I}(t) }{\int\limits_{-\infty}^{\infty}dt \mathcal{W}_{I}(t)}};
\label{5}
\end{equation}
while similar expressions can be used for calculating the total (over the period of the pulse) transmission $\bar {\mathcal{T}}$, and reflection 
$\bar {\mathcal{R}}$.

Our starting point is the ``rate'' equation 
\begin{equation}
\frac{d}{dt}T\left( t\right) =\frac{1}{C}\left(\mathcal{A}(T)\mathcal{W}_{I}\left( t\right)+\kappa (T_0-T)\right),  
\label{heat1}
\end{equation}
that describes the heating rate of the defect layer. Above, $C$ is the heat capacity, $\mathcal{W}_{I}\left( t\right)\equiv\left|
\mathcal{E}_{I}(t)\right|^{2}=\left|\int d\omega E(\omega) \exp(i\omega t)d\omega\right|^2$ is the incident light intensity, 
and $\kappa$ is the thermal conductance of the defect layer. The first term in Eq. (\ref{heat1}) describes the heating process 
of the lossy layer while the second one corresponds to heat dissipation from the defect layer to the mirror (if any) or to the 
air. To further simplify our calculations, we assume that the temperature changes are within a domain where both thermal 
conductance and heat capacity are constants and independent of temperature changes.

Substitution of the absorption coefficient from Eq. (\ref{Afin}) into Eq. (\ref{heat1}) leads us to the following equation 
\begin{equation}
\frac{d}{dt}T\left( t\right) =\frac{1}{C} \left(\omega \varepsilon_{\mathrm{d}}^{\prime\prime}(T) \rho(T) \mathcal{W}_{I}
(t) +\kappa (T_0-T)\right)
\label{heat4}
\end{equation}
which expresses the temporal behavior of the temperature $T(t)$ in terms of the given profile $\mathcal{W}_{I}(t)$ of the 
incident pulse. Everything else, e.g. $\epsilon_{\mathrm{d}}^{\prime \prime }(t)$, $\mathcal{A}(t)$, $\mathcal{T}(t)$ and 
$\mathcal{R}(t)$, can be directly and explicitly expressed in terms of $T(t)$.

In case that $\kappa=0$, one can further show that the outcomes can be written in terms of the total incident energy 
$U_f=\int_0^{t_f} \mathcal{W}_I(t) dt$. Furthermore, using Eq. (\ref{heat4}) we get that $T_f=\int_0^{U_f}\mathcal{A}(U) 
dU/C$. The associated total absorption is $\bar{\mathcal{A}} = \left(\int_0^{U_f} \mathcal{A}(U) dU\right)/U_f$, while 
similar expressions can be derived for the other transport characteristics.

In Fig. \ref{fig2} we report the outcomes of a direct integration of Eq. (\ref{heat4}) for the case of $\kappa=0$. In Fig. 
\ref{fig2}a we report the temporal behavior of permittivity $\epsilon_{d}^{\prime \prime}$ as a function of the pulse 
duration $t_f$. Notice that for train pulses the pulse duration $t_f$ is directly analogous of the total incident energy $U_f$. We
will therefore alternate, in our presentation below, the dependence of $\epsilon_{\mathrm{d}}^{\prime \prime}$, 
$\mathcal{T,R,A}$ from the pulse duration with the (more natural parameter for an energy limiter) total
incident energy of the pulse.

Originally $\epsilon_{d}^{\prime \prime}$ is essentially unaffected by the
incident energy and the same is true for the resonance mechanism (via the
defect mode) that is responsible for high transmittance in the absence of
losses. In this domain $\mathcal{T}\approx 1$, $\mathcal{R}\approx 0$ while
there is a slow increase of the absorption $\mathcal{A}$, as it can be seen
from Fig. \ref{fig2}b (solid lines). Once the incident energy (pulse
duration time) exceeds some critical value, there is a rather abrupt
increase in $\epsilon_{\mathrm{d}}^{ \prime \prime}$ which results to the
destruction of the resonance mode. Subsequently, the incident energy does
not resonate into the structure, leading to a decaying absorption $\mathcal{A}\approx 0$, while the same is true for 
the transmittance $\mathcal{T}\approx 0$. At the same time, there is a noticeable growth of the reflectance which 
becomes approximately equal to unity $\mathcal{R}\approx 1$. For comparison we also plot at the same figure the 
results of the stand-alone layer. We find that for large incident energies (pulse durations 
$t_f$) the absorption $\mathcal{A}(t)$ is higher by more than two orders of
magnitude as compared to the case of reflective energy limiter.

We have also performed the same analysis for the case where the thermal
conductance $\kappa $ is different from zero. In Fig. \ref{fig3} we report
the results of the numerical integration of Eq. (\ref{heat4}) in the
presence of thermal conductivity. For long pulse duration we find a
steady state behavior of the transport characteristics of
the reflective energy limiter. The physical nature of the steady-state
regime is quite obvious. It corresponds to the situation when the heat
released in the defect layer is completely carried away by thermal
conductivity. At this point, the temperature of the defect layer stabilizes
and the time derivative $dT(t)/dt$ in Eqs. (\ref{heat1},\ref{heat4})
vanishes. The latter condition determines the steady-state values of the
defect layer temperature as a function of the incident light amplitude. 
In this limiting case our structure acts as a power limiter.
For comparison, the results of the stand-alone lossy layer are also reported in 
this figure. We find that in the steady-state regime our structure 
performs superbly resulting in absorption values which are more than two 
orders of magnitude smaller than the onces achieved by the stand-alone lossy layer.


\section{Conclusions}

\label{conclusions}

At infrared and optical frequencies, the reflectivity of known uniform
materials is well below 90\%, especially so when the incident light
intensity is dangerously high. So, if we want to build a highly reflective
optical limiter, we have to rely on photonic structures which would support
some kind of low-intensity resonant transmission via slow or localized modes
at photonic band-gap frequencies. If the incident light intensity increases,
the respective localized mode must disappear, and the entire photonic
structure will behave as a simple Bragg reflector. Here we considered the
so-called "dissipative" mechanism of the localized mode suppression. At first
glance, it seems counterintuitive, because the high reflectivity and low
absorption are caused by the increase in the loss tangent of the defect
layer in Fig. \ref{fig1}. A qualitative explanation for such a phenomenon is
that the large value of  $\epsilon ^{\prime \prime }$ in the defect layer
results in decoupling of the left and the right Bragg reflectors in Fig. \ref{fig1}. Of course, there might be other ways to suppress resonant
transmittance when the incident light intensity, or the total energy of the
pulse, grow dangerously high. Still, the presented "dissipative" mechanism
seems simple and practical.

\textit{Acknowledgments -} This work is sponsored by the Air Force Office of
Scientific Research LRIR09RY04COR and by an AFOSR MURI grant
FA9550-14-1-0037.



\begin{thebibliography}{99}
\bibitem{notefield} We stress that in the case that both Bragg gratings on
the left and right of the defect layer are finite, both evanescent
contributions are present in either half-space, although, only one if them
is dominant on either side. Furthermore one can show that the presence of
both evanescent contributions on either side of the defect layer can provide
an energy flux and, hence, a non-zero transmittance.
\bibitem{notetrain} We have checked that the same qualitative behaviour is
obtained for other pulse shapes as well.
\end{thebibliography}
\end{document}